\documentclass{ifacconf}  
\usepackage{natbib}        

\pdfoutput=1
\usepackage{graphicx}      
\usepackage{grffile}       
\usepackage{booktabs}       
\usepackage{xparse}
\usepackage{mathbbol}
\usepackage{blindtext}
\usepackage[T1]{fontenc}
\usepackage{tikz}
\usetikzlibrary {graphs}
\usetikzlibrary{arrows.meta}

\definecolor{supelecRed}{RGB}{120,30,56}
\definecolor{mpc_blue}{RGB}{39,154,216}
\definecolor{mpc_green}{RGB}{98, 160, 98}
\definecolor{mpc_grey}{RGB}{235, 235, 235}
\definecolor{mpc_orange}{RGB}{247, 153, 68}
\definecolor{mpc_yellow}{RGB}{245, 235, 103}
\usetikzlibrary{calc,shapes,positioning}
\usetikzlibrary{overlay-beamer-styles}
 \usetikzlibrary{plotmarks}
  \usetikzlibrary{arrows.meta}

\usepackage{tikzscale}
\usepackage{bm}
\usepackage{ifthen}
\usepackage[ruled,noend,algo2e]{algorithm2e}
\SetKwRepeat{Do}{do}{while}

\usepackage{color}
\newcommand{\no}[1]{}

\makeatletter
\def\comments{
  \usepackage{geometry}
  \newgeometry{
    textwidth=8cm,
    hoffset=-1.5in,
    bottom=0.41in,
    vscale=.5,
    top=0.41in,
    footskip=1cm,
  }
\@TwoColumnfalse
\@twocolumnfalse
}
\makeatother

\newif\ifdebug%
\newcommand{\draft}{\debugtrue}
\newcommand{\final}{\debugfalse}

\makeatletter
\@ifpackageloaded{pdfcomment}
{

}
{

}
\makeatother

\newtheorem{remark}{Remark}
\newtheorem{assumption}{Assumption}
\newtheorem{challenge}{Challenge}

\graphicspath{{img/}}

\newcommand{\R}{\mathbb{R}}

\newcommand{\T}{^{\mathrm{T}}}
\newcommand{\1}{\mathbf{1}}
\newcommand{\0}{\mathbf{0}}

\newcommand{\norm}[1]{\lVert#1\rVert}

\newcommand{\set}[1]{\mathcal{#1}}
\newcommand{\p}{^{(p)}}
\newcommand{\pplusone}{^{(p+1)}}
\renewcommand{\vec}[1]{\boldsymbol{#1}}
\newcommand{\random}[1]{\underline{#1}}
\newcommand{\randomvec}[1]{{\underline{\vec{#1}}}}
\newcommand{\probability}[1]{\mathbb{P}(#1)}

\newcommand{\expectation}[2][]{\mathbb{E}_{#1}\left[#2\right]}
\newcommand{\indicator}[1]{\mathbb{1}_{\{#1\}}}
\newcommand{\until}{\mathbin{:}}

\newcommand{\setbuild}[2]{\{#1\mid#2\}}

\newcommand{\hadamard}[2]{#1\odot #2}
\newcommand{\kron}[2]{#1\mathbin{\otimes}#2}
\newcommand{\vectorize}[1]{\mathrm{vec} (#1)}
\newcommand{\symmetric}{\mathbb{S}}
\newcommand{\semidefpos}{\mathbb{S}_{+}}
\newcommand{\defpos}{\mathbb{S}_{++}}

\newcommand{\pseudoinv}[1]{{#1}^{\dagger}}

\usepackage{amsmath}
\usepackage{mathtools}
\usepackage{amssymb}

\DeclareMathOperator{\diag}{diag}
\DeclareMathOperator{\Proj}{Proj}

\newcommand{\nsubsystems}{M}
\newcommand{\umax}{\vec{u}_{\mathrm{\max}}}
\newcommand{\predhorz}{N_{p}}

\NewDocumentCommand \mpcvec { s m o o o } {%
  \IfBooleanTF{#1}{
    \def\optim{^\star}
  }{
    \def\optim{}
  }
  \IfValueTF{#5}{
    \vec{#2}_{#3}\optim{}[#4|#5]
  }{
    \IfValueTF{#4}{
      \vec{#2}_{#3}\optim{}[#4]
    }
    {
    \IfValueTF{#3}{
      \vec{#2}_i\optim{}[#3]
    }
    {
      \vec{#2}_i\optim{}[k]
    }
    }
  }
}

\NewDocumentCommand \mpcval { s m o o o } {%
  \IfBooleanTF{#1}{
    \def\optim{^\star}
  }{
    \def\optim{}
  }
  \IfValueTF{#5}{
    {#2}_{#3}\optim{}[#4|#5]
  }{
    \IfValueTF{#4}{
      {#2}_{#3}\optim{}[#4]
    }
    {
    \IfValueTF{#3}{
      {#2}_i\optim{}[#3]
    }
    {
      {#2}_i\optim{}[k]
    }
    }
  }
}

\newcommand{\optuikk}{\mpcvec*{u}[i][k][k]}

\newcommand{\globobj}{\mpcval{J}[][k]}
\newcommand{\optglobobj}{\mpcval*{J}[][k]}

\newcommand{\obji}{\mpcval{J}[i][k]}

\newcommand{\xik}{\mpcvec{x}}
\newcommand{\fik}{\mpcvec{f}}
\newcommand{\uik}{\mpcvec{u}}
\newcommand{\uiseq}{\mpcvec{u}[i][k:k+\predhorz-1][k]}
\newcommand{\optuiseq}{\mpcvec*{u}[i][k:k+\predhorz-1][k]}

\newcommand{\useq}{\mpcvec{u}[ ][k:k+\predhorz-1][k]}

\newcommand{\Uik}{\mpcvec{U}}

\newcommand{\optuncUik}{\mpcvec*{\mathring{U}}}

\newcommand{\vik}{\mpcvec{v}}
\newcommand{\wik}{\mpcvec{w}}

\newcommand{\Wik}{\mpcvec{W}}

\newcommand{\qik}{\mpcvec{q}}
\newcommand{\qiseq}{\mpcvec{q}[i][k:k+\predhorz-1][k]}
\newcommand{\thetaik}{\mpcvec{\theta}}

\newcommand{\thetai}{\vec{\theta}_i}
\newcommand{\optthetai}{\vec{\theta}_i^{\star}}

\newcommand{\dik}{\mpcvec{d}}
\newcommand{\diseq}{\mpcvec{d}[i][k:k+\predhorz-1][k]}
\newcommand{\lambdaik}{\mpcvec{\lambda}}
\newcommand{\lambdai}{\vec{\lambda}_i}
\newcommand{\lambdaicheat}{\tilde{\vec{\lambda}}_i}
\newcommand{\optlambdai}{\vec{\lambda}_i^{\star}}

\newcommand{\Tik}{\mpcval{T}}

\usepackage[acronym]{glossaries}%

\newcommand{\acrSing}[3]{\newacronym{#1}{#2}{#3}
  \expandafter\newcommand\csname #1\endcsname{\gls{#1}}}

\newcommand{\acrPl}[5]{
  \newacronym[plural=#4,firstplural=#5 (#4)]{#1}{#2}{#3}
  \expandafter\newcommand\csname #1\endcsname{\gls{#1}}
  \expandafter\newcommand\csname #4\endcsname{\glspl{#1}}
}
\newcommand{\acr}[5][4=,5=]{
  \ifthenelse{\equal{#5}{}}
  {
    \acrSing{#1}{#2}{#3}
  }
  {
    \acrPl{#1}{#2}{#3}{#4}{#5}
  }
}

\acrSing{mpc}{MPC}{Model-based Predictive Control}
\acrSing{qp}{QP}{\emph{quadratic program}}
\acrSing{dmpc}{dMPC}{distributed Model-based Predictive Control}
\acrSing{pwa}{PWA}{Piecewise Affine}
\acrSing{EM}{EM}{Expectation Maximization}
\acrSing{cps}{CPS}{cyber-physical systems}
\acrSing{fdi}{FDI}{false data injection}
\acrSing{rhs}{RHS}{receding horizon strategy}

\draft
\final

\begin{document}

\begin{frontmatter}
\title{\LARGE \bf
  Expectation-Maximization Based Defense Mechanism for Distributed Model Predictive Control
}

\author[First]{Rafael Accácio Nogueira}
\author[First]{Romain Bourdais}
\author[First]{Simon Leglaive}
\author[First]{Hervé Guéguen}
\address[First]{IETR-CentraleSupélec, 35510 Cesson-Sévigné, Ille-et-Vilaine, France\\
{\tt\small \{rafael-accacio.nogueira, romain.bourdais, simon.leglaive, herve.gueguen\}
@centralesupelec.fr}}




\begin{abstract}
  Controlling large-scale systems sometimes requires decentralized computation.
  Communication among agents is crucial to achieving consensus and optimal global behavior.
  These negotiation mechanisms are sensitive to attacks on those exchanges.
  This paper proposes an algorithm based on Expectation Maximization to mitigate the effects of attacks in a resource allocation based distributed model predictive control.
  The performance is assessed through an academic example of the temperature control of multiple rooms under input power constraints.
\end{abstract}

\begin{keyword}
Distributed constrained control and MPC;
Cyber security in networked control systems;
Decentralized Control and Large-Scale Systems
\end{keyword}

\end{frontmatter}

\section{INTRODUCTION}
The information collected, processed, and transmitted within \cps{} impacts physical systems whose proper functioning is essential for society, like energy networks, industrial processes, water distribution, and others.
Attacks in those critical infrastructures, such as Stuxnet, increased awareness for cyber-security of \cps{}.

Given these structures' complexity, scale, and geographic distribution, it is necessary to adopt decentralized control structures rather than monolithic ones.
For these systems, a set of controllers are designed to collectively ensure the safe and efficient operation of the facilities.

This operation is done cooperatively, where the controllers exchange information.
This approach can be used, i.e., for the voltage regulation of electrical networks, the production-consumption balance in these same networks, water distribution, and many other applications.
Multi-agent systems~\citep{KantamneniEtAl2015} and \dmpc{}~\citep{MaestreEtAl2014} are standard approaches to design these controllers.

A few articles have focused on safety in \dmpc{} frameworks, presenting vulnerabilities of different frameworks, such as Lagrange-based decomposition~\citep{VelardeEtAl2017}, Jacobi-Gauss decomposition~\citep{ChanfreutEtAl2018}, and primal decomposition~\citep{NogueiraEtAl2021}.

The authors in these works highlight the complexity of the problems. A modification of a local goal, the hacking of communication, or the failure of one of the agents are events that disrupt global behavior.
If one of these agents is attacked, it can result in the violent destruction of the corrupted element (perhaps the whole system), or more subtly, it can lead to a deviant behavior that is more difficult to detect.

Some strategies are presented to mitigate the effects of such an event.
For instance, dismissing extreme values as in~\cite{VelardeEtAl2017} or
using secure scenarios based on reliable historical data~\cite{MaestreEtAl2021}.

\cite{NogueiraEtAl2021} proposed a safe algorithm based on data reconstruction to mitigate the effects of a \fdi{} in the communication between a coordinator and local constraint-free agents, which were bound by global equality constraints.

In this paper, we propose an extension of their approach by including local constraints and changing the global constraints.
This extension profoundly changes the complexity, as the exchanges between the agents and the coordinator are no longer characterized by affine functions but by \pwa{} functions.
This fact leads not only to a combinatorial explosion but also to a parametric identification challenge.
To overcome this problem, we propose using a learning method based on the \EM{} algorithm~\citep{DempsterEtAl1977}, which allows us to estimate the corruption mechanism and correct it if necessary.

This paper is organized as follows.
First, Section~\ref{sec:PS} introduces the primal decomposition-based \dmpc{} and its vulnerabilities.
Then in Section~\ref{sec:TSM}, we study the local problems and propose a detection and mitigation mechanism based on the Expectation Maximization algorithm.
Moreover, in Section~\ref{sec:App_dMPC}, we illustrate the algorithm with an academic simulation and assess its performance.
Finally, in Section~\ref{sec:CC}, we conclude with an outlook of future works.

\emph{Notation:}
In this paper, $\norm{\cdot}$, $\norm{\cdot}_{F}$, and ${\norm{\vec{v}}_{Y}=\norm{Y^{\frac{1}{2}}\vec{v}}}$ represent the $\ell_{2}$, Frobenius and weighted norms.
$\Proj^{\set{T}}(\cdot)$ is the Euclidean projection onto $\set{T}$.
$\kron{}{}$ ($\hadamard{}{}$) is the Kronecker (Hadamard) product.
$\vectorize{A}$ vectorizes matrix $A$.
${n\until i\until j}$ is a row vector builder with elements $\{n,n+i,\dots,n+mi\}$, where ${m=\mathrm{truncate}(\frac{j-n}{i})}$, and ${n\until j}$ means ${i=1}$.
${\0_{m}=0_{m\times 1}}$ (${\1_{m}=1_{m\times 1}}$), where
$0_{m\times n}$ ($1_{m\times n}$) is a $0$ ($1$) filled \mbox{$m$-by-$n$} matrix.
$\indicator{x}$ is the indicator function returning $1$ if $x$ is true and $0$ otherwise.
$\symmetric^{n}$, and $\defpos^{n}$ ($\semidefpos^{n}$) are symmetric and positive (semi-)definite matrices of size $n$.
${A^{\dagger}={{(A^{T}A)}^{-1}A^{T}}}$.
A vector $\vec{v}_{i}$, corresponds to the $i$-th agent, and these vectors can be stacked in a vector $\vec{v}$.
$\expectation{\random{x}}$ is the expected value of random variable $\random{x}$, and $\probability{A \mid B}$ is the conditional probability of $A$ given condition $B$.
${\diag(A_{1},\dots,A_{N})}$ corresponds to a block diagonal matrix.

\section{Problem Statement}\label{sec:PS}
In this section, we present a \dmpc{} algorithm based on the primal decomposition \citep{BoydEtAl2015} and a vulnerability of the decomposition, which can be exploited.

\subsection{Model Predictive Control}\label{ssec:MPC}
Consider a system composed of discrete-time linear time-invariant agents ${i\in\set{M}=\{1\until \nsubsystems\}}$, modeled by
\begin{equation}
\begin{matrix}
  \label{eq:systems}
\vec{x}_{i}[k+1]&=&A_{i}\xik &+& B_{i}\uik\\
\vec{y}_{i}[k+1]&=&C_{i}\xik &&
\end{matrix},
\end{equation}
with ${\vec{x}_{i}[k]\in R^{n_{x}}}$, ${\vec{u}_{i}[k]\in R^{n_{u}}}$ and ${\vec{y}_{i}[k]\in R^{n_{y}}}$.

Each agent's input $\uik$ is constrained by
\begin{equation}
  \label{eq:u_local_constraints}
  \0_{n_{u}}\preceq\uik\preceq\vec{u}_{\max},
\end{equation}
with ${\umax\in R^{n_{u}}}$.
The agents are coupled by global input constraints with weighting matrices ${\Gamma_{i}\in\semidefpos^{n_{u}}}$:
\begin{equation}
  \label{eq:global_constraint}
  \sum_{i\in\set{M}}\Gamma_{i}\uik\preceq\umax.
\end{equation}

The overall system is controlled by a \mpc{}, which
computes the optimal input for a finite prediction horizon ${\set{H}=\{1\until \predhorz\}}$ by solving the following problem:
\begin{equation}\label{eq:GOP}
\resizebox{.88\columnwidth}{!}{$
\begin{matrix}
\underset{\useq}{\mathrm{minimize}}&\resizebox{0.35\textwidth}{!}{$\overbrace{\sum\limits_{i\in\set{M}} \overbrace{\sum_{j\in\set{H}}\norm{\mpcvec{v}[i][k+j][k]}^{2}_{Q_i}+\norm{\mpcvec{u}[i][k+j-1][k]}^{2}_{R_i}}^{\textstyle{} \obji}}^{\textstyle{} \globobj}$}\\
\mathrm{subject~ to}&~\eqref{eq:systems},\eqref{eq:u_local_constraints}\ \mathrm{and}\ \eqref{eq:global_constraint}
\left\}\small
\begin{aligned}
  &\forall i\in\set{M}\\
  &\forall j\in\set{H}
\end{aligned}\right.,

\end{matrix}
  $}
\end{equation}
where ${Q_{i}\in\semidefpos}$, ${R_{i}\in\defpos}$, and $\vik$ is a control objective.
$\optglobobj$ denotes the optimal value of the objective function for the optimal control sequences ${\optuiseq}$.
The problem in~\eqref{eq:GOP} is solved at each time $k$, and the ${\optuikk}$ are applied in their respective subsystem $i$, following a \rhs{}.

Since the prediction horizon and the number of agents can be large, solving~\eqref{eq:GOP} at each time $k$ can be rather challenging due to computational costs.
Another issue that can arise is the fact that the complete information ${\set{I}_i=\{A_i,B_i,C_i,Q_i,R_i,\Gamma_i,\vec{v}_i\}}$ from all the subsystems is needed to solve the problem,
which can be viewed negatively from a confidentiality point of view.

So, we use the \emph{primal decomposition} method, sometimes called \emph{quantity decomposition} or \emph{resource allocation}, with which we can profit from the parallelism, while avoiding the use of the complete information $\set{I}_{i}$.

\subsection{Primal Decomposition based dMPC}\label{ssec:dMPC}

The technique decomposes the monolithic \mpc{} in~\eqref{eq:GOP} into $M$ modified \mpc{} problems~\eqref{eq:DOP_local} (solvable in parallel by each subsystem) called \emph{local problems}, and a \emph{master problem}~\eqref{eq:DOP_master}, which is equivalent to the global problem and is solved by a coordinator.
\begin{subequations}
  \begin{equation}
    \left.
      \small
        J_{i}^{\star}(\thetaik)=
        \begin{matrix}
        \underset{\uiseq}{\mathrm{minimize}}&\obji\\
        \mathrm{subject~ to} &\eqref{eq:systems}\ \&\ \eqref{eq:u_local_constraints}\\
        &\Gamma_{i}\uik\preceq\qik:\dik\\
      \end{matrix}
    \right\}
    \small
    \begin{aligned}
      &\forall i\in\set{M}\\
      &\forall j\in\set{H}
    \end{aligned}
    \label{eq:DOP_local}
  \end{equation}

  \begin{equation}
    \small
    \begin{aligned}
      J^{\star}=
      \begin{matrix}
        \underset{\qiseq}{\mathrm{minimize}} &\sum_{i\in\set{M}} J^{\star}_i(\qik)\\
        \mathrm{subject~ to} &
        \left.
          \begin{aligned}
            \quad \sum_{i\in\set{M}}\qik\preceq\umax\\
            \quad \qik[k]\succeq\0_{n_{u}},       \forall i\in\set{M}
          \end{aligned}
        \right\}
        \begin{aligned}
          \forall j\in\set{H}
        \end{aligned}
      \end{matrix}
    \end{aligned}
    \label{eq:DOP_master}
  \end{equation}
\end{subequations}
The modified \mpc{} problems~\eqref{eq:DOP_local} are created by exchanging the $\vec{u}_{\max}$ in~\eqref{eq:global_constraint} by vectors $\qik$, which correspond to the quantity of the total resource $\vec{u}_{\max}$ allocated to agent $i$ in time $k$ for each prediction $j$, thus the names \emph{resource allocation} and \emph{quantity decomposition}.
This new set of constraints have associated dual variables ${\dik\succeq\0_{n_{u}}}$.
The sequences ${\qiseq}$ and ${\diseq}$ can be aggregated in vectors ${\thetaik=\kron{\1_{\predhorz}}{\qik}}$, and ${\lambdaik=\kron{\1_{\predhorz}}{\dik}}$.

The \emph{master problem} finds the optimal allocation sequences $\optthetai$ by using a projected sub-gradient method:
\begin{equation}
  \label{eq:projectedSubgradient}
\vec{\theta}[k]\pplusone=\Proj^{\set{S}}(\vec{\theta}[k]\p-\rho\p\vec{g}[k]\p),
\end{equation}
where ${\set{S} = \setbuild{\vec{\theta}}{I_{c}^{M}\vec{\theta}\preceq \vec{U}_{\max}\ \&\ \vec{\theta}\succeq\0 }}$, being ${c=\predhorz n_{u}}$, ${I_{c}^{M}=\kron{\1_{M}}{I_{c}}}$, ${\vec{U}_{\max}=\kron{\1_{\predhorz}}{\vec{u}_{\max}}}$, $(p)$ is a given step in the iterative process and $\vec{g}\p[k]$ is a sub-gradient of the objective function of problem in~\eqref{eq:DOP_master} in step $(p)$.

From~\cite{BoydEtAl2015}, it is known that $-\lambdaik$ are sub-gradients of objective $J[k]$.
Plugging it in $\vec{g}[k]\p$ we have
\begin{equation}
  \label{eq:thetaNegot}
\vec{\theta}[k]\pplusone=\Proj^{\set{S}}(\vec{\theta}[k]\p+\rho\p\vec{\lambda}[k]\p),
\end{equation}
\newcommand{\negotiation}{\emph{negotiation}}
which henceforth is referred to as the \negotiation.

Once the \negotiation{} for a given time $k$ converges, the $\optuikk$ found in the last step of the \negotiation{} are applied in their corresponding agent and then the \rhs{} is followed.

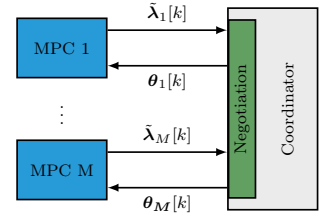
\begin{figure}[h]
  \centering
    \resizebox{.47\columnwidth}{!}{
          \begin{tikzpicture}[font=\small,thick,node distance=.0cm and .5cm,
          mpcSmall/.style={
            rectangle,
            align=center,
            fill=supelecRed!10,
            minimum width=1.5cm,
            minimum height=1cm},
          coordinator/.style={
            rectangle,
            align=center,
            fill=mpc_grey,
            minimum height=3.4cm,
            minimum width=1.5cm,
          },
          superv/.style={
            rectangle,
            minimum height=1.4cm,
            minimum width=.3cm,
            fill=mpc_yellow,
          },
          negotiation/.style={
            rectangle,
            minimum height=3.0cm,
            minimum width=.45cm,
            fill=mpc_green,
          },
          ]

          \node[draw,
          mpcSmall,
          fill=mpc_blue,
          ] (block1) {MPC 1};

          \node[draw,
          mpcSmall,
          fill=none,
          draw=none,
          below=of block1,
          ] (mult) {\vdots};

          \node[draw,
          mpcSmall,
          fill=mpc_blue,
          below=of mult,
          ] (blockM) {MPC M};

          \node[draw,
          coordinator,
          right=2cm of mult,
          ] (coordinator) {};
          \node at ($(coordinator)+(.25,0)$) {\rotatebox{90}{Coordinator}};

          \node[draw,negotiation] (negot) at ({$(coordinator.center)+(0,0)$} -|{$(coordinator.west)+(0.25,)$}) {};
          \node  at (negot) {\rotatebox{90}{Negotiation}};

          \draw[-latex,thick] (block1.east)+(0,.3) -- ( {$(block1.east)+(0,.3)$} -| coordinator.west) node [above,midway] {$\tilde{\vec{\lambda}}_{1}[k]$};
          \draw[latex-,thick] (block1.east)+(0,-.3) -- ( {$(block1.east)+(0,-.3)$} -| coordinator.west) node [below,midway] {$\vec{\theta}_{1}[k]$};

          \draw[-latex,thick] (blockM.east)+(0,.3) -- ( {$(blockM.east)+(0,.3)$} -| coordinator.west) node [above,midway] {$\tilde{\vec{\lambda}}_{M}[k]$};
          \draw[latex-,thick] (blockM.east)+(0,-.3) -- ( {$(blockM.east)+(0,-.3)$} -| coordinator.west) node [below,midway] {$\vec{\theta_{M}}[k]$};

        \end{tikzpicture}
        }
  \caption{Exchange between coordinator and agents.}\label{fig:dmpc_graph}
\end{figure}

The algorithm to find the optimal $\optthetai[k]$ can be summarized in Algorithm~\ref{alg:quantityAlg}, and the exchange between coordinator and agents can be seen in Fig.~\ref{fig:dmpc_graph}.

\begin{algorithm2e}[h]
  \DontPrintSemicolon%
  $p:=0$\;
  Coordinator initializes $\vec{\theta}\p$ \;
  \Repeat{$\|\vec{\theta}^{(p)} -\vec{\theta}^{(p-1)}\|\leq\epsilon$}{
  Subsystems solve~\eqref{eq:DOP_local}, and send $\optlambdai (\thetai\p)$\;
  Coordinator updates allocations~\eqref{eq:thetaNegot}\;
  $p:=p+1$
}
 \caption{Quantity decomposition negotiation.}\label{alg:quantityAlg}
\end{algorithm2e}

One can observe that each agent only sends $\lambdaik$ instead of using $\mathcal{I}_{i}$.
An interpretation of $\lambdaik$ is the dissatisfaction of agent $i$ with the resource $\thetaik$ allocated for it, where ${\lambdaik=\0_{c}}$ means total satisfaction.
The coordinator trusts in the authenticity of the $\vec{\lambda}_{i}[k]$ received to update the allocations. However, if false data is injected, the \negotiation{} can be driven by a malicious agent, taking advantage of this trust to favor itself, harm others or even destabilize the \negotiation{}, as shown in~\cite{NogueiraEtAl2021}.

Our main objective is to mitigate the effects of a given configuration where agents send untrustworthy
\begin{equation*}
\lambdaicheat=\gamma_{i}(\lambdai).
\end{equation*}

\section{Towards a safe dMPC}\label{sec:TSM}

Here we focus on the local problems in~\eqref{eq:DOP_local}, and study how their structure can contribute for a safe dMPC algorithm.

\subsection{Local Problems --- Formal Analysis}\label{ssec:FA}

The \emph{local problems}~\eqref{eq:DOP_local} can be rewritten in a equivalent (same solution) constrained \qp{} form:
  \begin{subequations}\label{eq:quadratic_case}
    \begin{align}
      \label{eq:4}
      \underset{\Uik}{\mathrm{minimize}}&&\small J_{i}(\vec{\theta}_{i})=&\frac{1}{2} \Uik^T H_i\Uik +{\fik}^T\Uik\\
      \mathrm{subject~ to}&&&\bar{\Gamma}_{i}\Uik\preceq\vec{\theta}_{i}[k]:\vec{\lambda}_{i}[k]\label{eq:constraints}\\
                                        &&&\Uik\succeq\0_{c},
    \end{align}
  \end{subequations}
where ${\Uik}$ stacks the input prediction sequences, and
${\bar{\Gamma}_{i}=\kron{I_{\predhorz}}{\Gamma_{i}}}$.
The values for $H_{i}$ and $\vec{f}_{i}[k]$ vary depending on the control objective $\vec{v}_{i}[k]$.
The approach presented in this paper works for linear control objectives such as ${\vec{v}_{i}[k]=\vec{y}_{i}[k]}$ and ${\vec{v}_{i}[k]=\vec{y}_{i}[k]-\vec{w}_{i}[k]}$.
The approach depends only on the fact that ${H_{i}\in\semidefpos}$ does not vary w.r.t. $k$, while $\fik$ does. Those facts are instrumental for the results in the following sections.

\no{
  We choose reference tracking to illustrate the approach. With this objective, we have
${\vik=\wik-C_{i}\xik}$, where $\wik$ is the output reference in time $k$, which yields
\begin{equation}
\small\begin{matrix*}[l]
 H_i&=&\mathcal{D}_i^T\bar{Q}_i\mathcal{D}_i+\bar{R}_i\\
\fik&=&\mathcal{D}_i^T\bar{Q}_i(\mathcal{M}_i\xik-\Wik),
\end{matrix*}
\label{eq:matrices}
\end{equation}
where $\Wik$ stacks the setpoint prediction sequence, which is supposed to be constant during the predicted horizon ${\Wik=\kron{\1_{\predhorz}}{\vec{w}_{i}[k]}}$ (no prior information).
${\mathcal{M}_{i}}$  and ${\mathcal{D}_{i}}$ are the prediction matrices of the \mpc{}.
${\bar{Q}_{i}=\kron{I_{\predhorz}}{Q_{i}}}$, and ${\bar{R}_{i}=\kron{I_{\predhorz}}{R_{i}}}$.
}

We can get an explicit solution for the dual variables $\vec{\lambda}_{i}[k]$, which are \acrfull{pwa} functions w.r.t. $\vec{\theta}_{i}[k]$:
\begin{equation}
\resizebox{.88\columnwidth}{!}{$
  \begin{aligned}\label{eq:lambdafuntheta}
    \lambdaik=
      -P_{i}^{n}\thetaik-\vec{s}_{i}^{n}[k]\text{, if}\ G^{n}[k]\thetaik \preceq \vec{b}^{n}[k],
  \end{aligned}
  $}
\end{equation}
with $n\in\{1\until N\}$.

The halfspaces defined by the pairs $(G^{n}[k],\vec{b}^{n}[k])$ represent the combinations of active constraints~\eqref{eq:constraints} for a given time $k$, which vary w.r.t. $\vec{v}_{i}[k]$.

The $P_{i}^{n}$ are constructed using $H_{i}$ and $\bar{\Gamma}_{i}$; and $\vec{s}_{i}^{n}$ are constructed using $H_{i}$, $\fik$ and $\bar{\Gamma}_{i}$. The specific construction depends on the active constraints in~\eqref{eq:constraints}.
For example, if all constraints are active, we have ${P_{i}^{\text{ac}}={(\bar{\Gamma}_{i}H_{i}^{-1}\bar{\Gamma}_{i}\T)}^{-1}}$ and ${\vec{s}_{i}^{\text{ac}}[k]=P_{i}^{\text{ac}}\bar{\Gamma}_{i}H_{i}^{-1}\fik}$.
Futhermore, if all constraints are inactive, we have ${P_{i}^{\text{in}}=0_{c\times c}}$ and ${\vec{s}_{i}^{\text{in}}=\0_{c}}$.

\begin{remark}\label{rmk:P_constant}
The $\vec{s}_{i}^{n}[k]$ depend on time $k$, while $P_{i}^{n}$ do not.
\end{remark}

\begin{challenge}\label{ch:zone_unknown}
  It is important to note that since $\vec{v}_{i}[k]$ is unknown by the coordinator, it cannot anticipate the partition of the space for each agent.
\end{challenge}
\begin{challenge}\label{ch:s_unknown}
 For the same reason, the values of $\vec{s}_{i}[k]$ are also unknown.
\end{challenge}

 As each constraint can be active or inactive, for a given group of $t$ constraints, we can have at most ${N=2^{t}}$ different combinations of active sets.

 \begin{assumption}
   We assume that none of the active/inactive constraints combinations are redundant (no linear dependency), nor make the optimization infeasible (no empty intersection).
   Thus we always have $N$ zones.
 \end{assumption}

\subsection{The Attack}\label{ssec:attack}
We suppose the malicious agent chooses a function $\gamma_{i}(\cdot)$ to use throughout a given time $k$.
\begin{assumption}
  $\gamma_{i}(\cdot)$ does not change during the \negotiation{} phase for a given time $k$.
\end{assumption}

\begin{assumption}
  The agent chooses a linear function
  \begin{equation}
\tilde{\lambdai}=\gamma_{i}(\vec{\lambda}_{i})=\Tik\vec{\lambda}_{i},
\end{equation}
  where $T$ is invertible.
\end{assumption}

Such attack function results in a \pwa{} solution for $\lambdaicheat[k]$:
\begin{equation}
  \resizebox{.88\columnwidth}{!}{$
    \label{eq:linear_cheating}
    \lambdaicheat[k]=
      -\tilde{P}_{i}^{n}[k]\thetaik-\tilde{\vec{s}}_{i}^{n}[k]\text{, if}\ G^{n}[k]\thetaik \preceq b^{n}[k],
    $}
\end{equation}
with $n\in\{1\until N\}$, $\tilde{P}_{i}^{n}[k]=\Tik P_{i}^{n}$ and $\tilde{\vec{s}}_{i}^{n}[k]=\Tik\vec{s}_{i}^{n}[k]$.
Observe that as the function is applied only in $\lambdaik$, it does not affect the zones' hyperplanes.

We fix the index for the zones where all constraints are active to $1$, now referenced to as zones $1$ or the $1$-zones.

\subsection{Detection and mitigation}\label{ssec:DM}
Supposing we have a sequence of $\thetaik$ in a zone $j$, we can estimate the parameters for this zone with
\begin{equation}
  \label{eq:lambdafuntheta_tilde}
\lambdaicheat[k]=\gamma_{i}(\lambdai(\thetaik))=-\widehat{\tilde{P}_{i}^{j}}[k]\thetaik-\widehat{\tilde{\vec{s}}_{i}^{j}}[k].
\end{equation}

\begin{assumption}\label{ass:Pnominal}
  The nominal value of $P_{i}^{j}$ for this given $j$-zone, denoted $\bar{P}_{i}^{j}$, is available from reliable attack-free historical data.
\end{assumption}

Since $P_{i}^{j}$ do not change w.r.t.\ time, we can detect a deviation from nominal behavior using ${E_{i}[k] =\|\widehat{\tilde{P}_{i}^{j}}[k]-\bar{P}_{i}^{j}\|_{F}}$.
Let ${\mathfrak{D}_{i}}$ be an indicator
\begin{equation}
  \label{eq:2}
  \mathfrak{D}_{i}=\indicator{E_{i}[k]\geq\epsilon_{P}},
\end{equation}
which detects the attack in agent $i$
if the disturbance $E_{i}[k]$ disrespects an arbitrary bound $\epsilon_{P}$.

If an attack is detected, we can estimate the inverse of $T_{i}(k)$ with
\begin{equation}
\widehat{{T_{i}(k)}^{-1}}=\bar{P}_{i}^{j}{\widehat{\tilde{P}_{i}^{j}}[k]}^{-1},
\end{equation}
if $\widehat{\tilde{P}_{i}^{j}}[k]$ is invertible, and $\widehat{\tilde{P}_{i}^{j}}[k]$ is only invertible when all constraints are active, that means, when ${j=1}$.
\begin{assumption}
  For each agent $i$, at every time $k$, there is a corresponding $1$-zone.
\end{assumption}
Moreover, from~\eqref{eq:lambdafuntheta}, we can reconstruct $\vec{\lambda}_{i}$:
\begin{equation}
  \label{eq:lambda_reconstruction}
  {\vec{\lambda}_{i}}_{\mathrm{rec}}=\widehat{{T_{i}[k]}^{-1}} \tilde{\vec{\lambda}_{i}}
\end{equation}

In order to estimate $\widehat{\tilde{P}_{i}^{1}}[k]$, we must have enough observed points in the $1$-zones, so
we propose to generate points surrounding arbitrary $\bar{\vec{\theta}_{i}}$ in the $1$-zones.

\begin{assumption}\label{ass:}
  Given the unconstrained solution of~\ref{eq:quadratic_case} ${ {\optuncUik}=-H_{i}^{-1}\fik}$, we suppose
$\bar{\Gamma}_{i}\optuncUik\succeq\0_{c}$ for all $k$, that means, the points $\thetai=\0_{c}$ are in the $1$-zones.
\end{assumption}

\glsreset{EM}
Unfortunately, since we do not know the hyperplanes separating the different zones, we do not know how close of $\bar{\vec{\theta}}_{i}=\0_{c}$ we need to generate our points.
So we generate points arbitrarily close to $\bar{\vec{\theta}}_{i}$ and use the \EM{} algorithm, which can potentially identify the parameters of all $N$ modes.
\subsection{Expectation Maximization}
As the method is the same for all agents and repeated each time $k$, we drop the subscript $i$ and the time dependency $[k]$ to simplify the notation.

Given a set of parameters $\set{P}$, a set of observable data $\set{B}$ and a set of non-observable data $\set{U}$, the main objective of the \EM{} algorithm is to find estimators of $\set{P}$ that maximize the log marginal likelihood of the observed data ${\ln\probability{\set{B};\set{P}}}$, for models with latent variables in $\set{U}$.

Since maximizing ${\ln\probability{\set{B};\set{P}}}$ does not have an analytical solution, the algorithm solves the optimization problem iteratively.
So, rather than finding the $\set{P}$ that maximizes ${\ln\probability{\set{B};\set{P}}}$, we find the $\set{P}$ that maximizes the expectation of the complete data log-likelihood $\ln\probability{\set{B},\set{U};\set{P}}$ w.r.t.
the posterior probabilities ${\probability{\set{U}|\set{B};\set{P}}}$
calculated using a given set of parameter estimates $\set{P}_{\mathrm{cur}}$.
These steps provide a convergence-guaranteed iterative algorithm ensuring a monotonic increase of the log-marginal likelihood at each iteration (Algorithm~\ref{alg:em}).

\SetKwBlock{Estep}{ E step:}{}
\SetKwBlock{Mstep}{ M step:}{}
\begin{algorithm2e}[h]
  \DontPrintSemicolon%
  Initialize parameters $\set{P}_{\mathrm{new}}$\;
  \Repeat{$\set{P}_{\mathrm{cur}}$ converges}{
    $\set{P}_{\mathrm{cur}}\gets\set{P}_{\mathrm{new}}$\;
    \Estep{
      Evaluate ${\probability{\set{U}|\set{B};\set{P}_{\mathrm{cur}}}}$\;
    }
    \Mstep{
      Reestimate parameters using:
      \begin{equation}
      \resizebox{.68\columnwidth}{!}{$
        \label{eq:m_step}
        \set{P}_{\mathrm{new}}=\arg\underset{\set{P}}{\max\!.}\ \expectation[{\probability{\set{U}|\set{B};\set{P}_{\mathrm{cur}}}}]{\ln\probability{\set{B},\set{U};\set{P}}}
        $}
      \end{equation}
    }
}
 \caption{Expectation Maximization}\label{alg:em}
\end{algorithm2e}

For a group of $O$ exchanges between an agent and the coordinator, we observe the input and response variables, identified as  ${\random{\vec{\theta}}_{o}}$ and ${\random{\vec{\lambda}}_{o}}$, ${o\in\set{O}=\{1\until O\}}$, which can be organized in ${\random{\Theta},\random{\Lambda}\in\R^{c\times O}}$. ${(\random{\Theta},\random{\Lambda})}$ is our set of observable data $\set{B}$.

As~\eqref{eq:linear_cheating} gives us a multidimensional \pwa{} function, we propose using a multidimensional expansion of the model referred to as \emph{mixture of linear regressions} in~\cite{FariaSoromenho2010} to map the relation between $\random{\Theta}$ and $\random{\Lambda}$. We will call the model \emph{mixture of affine regressions}, since our regressors have linear and constant terms:
\begin{equation}\label{eq:linear_cheating_random}
  \randomvec{\lambda}_{o}=
    -\tilde{P}^{z}\randomvec{\theta}_{o}-\tilde{\vec{s}}^{z}\text{, if in zone } z,\\
\end{equation}
with  ${z\in\set{Z}=\{1\until Z\}}$.
\remark{Observe that the indices $z$, do not necessarily correspond to the original indices $n$.}

As \eqref{eq:linear_cheating_random} is a \pwa{} function whose modes depend on the unknown zone ${z\in\set{Z}}$,
we associate to each couple $(    \randomvec{\lambda}_{o}, \randomvec{\theta}_{o})$ a latent unobserved random variable ${\random{z}_{o}}$ that indicates in which zone in $\set{Z}$ the observable variables were obtained.
These variables are organized in ${\random{Z}\in\R^{1\times O}}$ which is our set of latent variables $\set{U}$.

The latent variable $\random{z}_{o}$ follows a categorical prior distribution, with associated probabilities ${\Pi=\{\pi^{z}|z\in\set{Z}\}}$ such
\[\probability{\random{z}_{o}=z}=\pi^{z} \in [0,1], \qquad \sum_{z\in\set{Z}} \pi^{z} = 1.
\]
As parameters to estimate, we have ${\set{P}=\setbuild{\set{P}^{z}}{z\in\set{Z}}}$, with ${\set{P}^{z}=(\tilde{P}^{z},\tilde{\vec{s}}^{z},\pi^{z})}$.

Since $\vec{\theta}$ is our input, we consider a non-informative improper probability density function~\citep{ChristensenEtAl2010}
\[
  \probability{\randomvec{\theta}_{o}} \,\propto \,1.
\]

Given the input and latent variables, the response variable $\random{\lambda}_{o}$ is modeled as a multivariate normal random variable with probability density function
\begin{equation}
  \label{eq:multivariate_gaussian}
\probability{\randomvec{\lambda}_{o}|\randomvec{\theta}_{o},\random{z}_{o}=z; \set{P}^{z}} = \mathcal{N}(\randomvec{\lambda}_{o};f(\randomvec{\theta}_{o};\set{P}^{z}),{\Sigma^{z}}),
\end{equation}
where, following~\eqref{eq:linear_cheating_random}, the mean vector is defined by
\({f(\randomvec{\theta}_{o};(P,\vec{s},\pi))=-{P}\randomvec{\theta}_{o}-{\vec{s}},}\)
and the covariance matrix ${\Sigma^{z}}$ tends to $0$.

\no{
This mixture of affine regression model corresponds to the following factorization of the complete-data (i.e., observed and latent data) likelihood, \cite{Bishop2006}:
\begin{align}\label{eq:completedataLikelihood}
  \probability{\random{\Theta},\random{\Lambda},\random{Z};\set{P}}= \prod_{\set{O}}\prod_{\set{Z}}\big[&\probability{\randomvec{\lambda}_{o}|\randomvec{\theta}_{o},\random{z}_{o}=z;\set{P}^{z}} \nonumber \\
& \times \probability{\random{z}_{o}=z}\probability{\randomvec{\theta}_{o}}\big]^{\indicator{\random{z}_{o}=z}},
\end{align}
}

We can calculate the posterior probabilities  $\zeta_{zo}(\set{P})=\probability{\random{z}_{o}=z|\randomvec{\lambda}_{o},\randomvec{\theta}_{o};\set{P}}$, also called \emph{responsibilities}:
\begin{align}
  \label{eq:responsibilities}
\zeta_{zo}(\set{P})&=\frac{\pi_{z}{\mathcal{N}(\randomvec{\lambda}_{o};f(\randomvec{\theta}_{o};\set{P}^{z}),{\Sigma^{z}})}}{\sum\limits_{j=1}^{Z}\pi_{j}\mathcal{N}(\randomvec{\lambda}_{o};f(\randomvec{\theta}_{o};\set{P}^{j}),{\Sigma^{j}})},
\end{align}
and then calculate the expectation of ${\ln\probability{\random{\Theta},\random{\Lambda},\random{Z};\set{P}}}$ with respect to ${\zeta_{zo}(\set{P}_{\mathrm{cur}})}$ \citep[Chapter 9]{Bishop2006}
\begin{align}
  \label{eq:completedataLogLikelihood_expectation}
\expectation[{\zeta_{zo}(\set{P}_{\mathrm{cur}})}]{\ln\probability{\random{\Theta},\random{\Lambda},\random{Z};\set{P}}}&= \sum_{o\in\set{O}}\sum_{z\in\set{Z}}  \zeta_{zo}(\set{P}_{\mathrm{cur}})\alpha_{zo},
\end{align}
where ${\alpha_{zo}=\ln{\pi_{z}}+\ln{\mathcal{N}(\randomvec{\lambda}_{o};f(\randomvec{\theta}_{o};\set{P}^{z}),{\Sigma^{z}})}}$.

\begin{remark}
The $\Sigma^{z}$ used in~\eqref{eq:responsibilities} are updated using a technique called \emph{Simulated annealing}~\citep{OzerovFevotte2010}, where they are initialized with arbitrarily significant values indicating the uncertainty of the parameters and it is reduced as the estimations converge.
\end{remark}

Here we introduce a variable ${\vec{\phi}^{z}={[\vectorize{\tilde{P}^{z}}^{T}\ {(\tilde{\vec{s}}^{z})}^{T} ]}^{T}}$.
We can find an optimal $\vec{\phi}^{z}$ for the problem in~\eqref{eq:m_step}, by
taking the gradients of~\eqref{eq:completedataLogLikelihood_expectation} with respect to vectors $\vec{\phi}^{z}$ and making them vanish.
Because of the multidimensional nature of the problem, some matrix operations are needed to synthesize the results.
After those operations, we have a matricial solution that yields the optimal estimates $\vec{\phi}^{z}_{\mathrm{new}}$:
\begin{equation}
  \label{eq:mstepestimation}
  \vec{\phi}^{z}_{\mathrm{new}}=\pseudoinv{(\Xi^{z}\random{\Omega})}\Xi^{z}\vectorize{\random{\Lambda}},
\end{equation}
where
${\random{\Omega}=[\hadamard{(\Upsilon \random{\Theta}\Delta)}{Y};G]}$,
with matrices
${\Upsilon=\kron{\1_{c}^{T}}{I_{c}}}$,
${\Delta=\kron{I_{O}}{\1_{c}^{T}}}$,
${Y=\kron{G}{\1_{c}}}$,
${G=\kron{\1_{O}^{T}}{I_{c}}}$,
and
\[{\Xi^{z}={\diag(\sqrt{{\zeta(z_{z1};\set{P}_{\mathrm{cur}})}}I_{c},\cdots,\sqrt{{\zeta(z_{zO};\set{P}_{\mathrm{cur}})}}I_{c})}}.\]
Doing the same for $\pi^{z}$ we get
\begin{equation*}
  \label{eq:3}
  \pi^{z}=\sum\limits_{o\in\set{O}}\tfrac{\zeta_{zo}(\set{P}_{\mathrm{cur}})}{O}.
\end{equation*}

As we can see,~\eqref{eq:mstepestimation} is the solution of a weighted Least-Squares, with responsibilities as weights, which adjust the contribution of all observations to the regression models.
We can see some similarities to the K-planes algorithm (see~\cite{BradleyMangasarian2000}), but \EM{} is more compromising.
Instead of affecting the observed data to a zone with 100\% of certainty (\emph{hard assignment}), \EM{} uses each zone's responsibilities (\emph{soft assignment}).

Once the estimates $\vec{\phi}_{z}^{\mathrm{new}}$ converge, we can reconstruct the estimates $\tilde{P}^{z}$ and $\tilde{\vec{s}}^{z}$, and use them in our mitigation scheme proposed in \S\ref{ssec:DM}.
Since the $z$ indices do not necessarily correspond to the indices $n$ in~\eqref{eq:lambdafuntheta} (due to initialization of $\set{P}$),
to find the $z$ that corresponds to ${n=1}$, we take the observation $o$ for $\bar{\vec{\theta}}=\0_{c}$, which we know belongs to the $1$-zone, and we find the most probable $z$ with
\begin{equation*}\label{eq:argmaxz}
\arg\underset{z}{\max\!.}\ {\zeta_{zo}(\set{P})}.
\end{equation*}

\begin{remark}
A discussion about the initialization and update of other parameters is beyond the scope of this article, so we refer the reader to~\cite{Bishop2006}.
\end{remark}

\subsection{Safe dMPC Algorithm}\label{ssec:safe_algo}
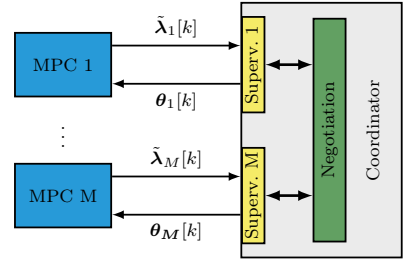
\begin{figure}[t]
  \centering
  \resizebox{.6\columnwidth}{!}{
          \begin{tikzpicture}[font=\small,thick,node distance=.0cm and .5cm,
          mpcSmall/.style={
            rectangle,
            align=center,
            fill=supelecRed!10,
            minimum width=1.5cm,
            minimum height=1cm},
          coordinator/.style={
            rectangle,
            align=center,
            fill=mpc_grey,
            minimum width=2.5cm,
            minimum height=4cm
          },
          superv/.style={
            rectangle,
            minimum height=1.5cm,
            minimum width=.35cm,
            fill=mpc_yellow,
          },
          negotiation/.style={
            rectangle,
            minimum height=3.5cm,
            minimum width=.5cm,
            fill=mpc_green,
          },
          ]

          \node[draw,
          mpcSmall,
          fill=mpc_blue,
          ] (block1) {MPC 1};

          \node[draw,
          mpcSmall,
          fill=none,
          draw=none,
          below=of block1,
          ] (mult) {\vdots};

          \node[draw,
          mpcSmall,
          fill=mpc_blue,
          below=of mult,
          ] (blockM) {MPC M};

          \node[draw,
          coordinator,
          right=2cm of mult,
          ] (coordinator) {};
          \node at ($(coordinator)+(.8,0)$) {\rotatebox{90}{Coordinator}};

          \node[draw,superv] (superv1) at ({$(block1)+(0,0)$} -|{$(coordinator.west)+(.2,)$}) {};
          \node  at (superv1) {\rotatebox{90}{Superv. 1}};

          \node[draw,superv] (supervM) at ({$(blockM)+(0,0)$} -|{$(coordinator.west)+(.2,)$}) {};
          \node  at (supervM) {\rotatebox{90}{Superv. M}};

          \node[draw,negotiation] (negot) at ({$(coordinator.center)+(0,0)$} -|{$(superv1.east)+(1.0,)$}) {};
          \node  at (negot) {\rotatebox{90}{Negotiation}};

          \draw[-latex,thick] (block1.east)+(0,.3) -- ( {$(block1.east)+(0,.3)$} -| coordinator.west) node [above,midway] {$\tilde{\vec{\lambda}}_{1}[k]$};
          \draw[latex-,thick] (block1.east)+(0,-.3) -- ( {$(block1.east)+(0,-.3)$} -| coordinator.west) node [below,midway] {$\vec{\theta}_{1}[k]$};

          \draw[latex-latex,very thick] (supervM.east) -- ( {$(supervM.east)$} -| negot.west) node [above,midway] {};
          \draw[latex-latex,very thick] (superv1.east) -- ( {$(superv1.east)$} -| negot.west) node [above,midway] {};

          \draw[-latex,thick] (blockM.east)+(0,.3) -- ( {$(blockM.east)+(0,.3)$} -| coordinator.west) node [above,midway] {$\tilde{\vec{\lambda}}_{M}[k]$};
          \draw[latex-,thick] (blockM.east)+(0,-.3) -- ( {$(blockM.east)+(0,-.3)$} -| coordinator.west) node [below,midway] {$\vec{\theta_{M}}[k]$};

        \end{tikzpicture}
        }
  \caption{Exchange between agents in secure \dmpc{}.}\label{fig:safe_dmpc_graph}
\end{figure}

Integrating the EM to our detection and mitigation mechanism we can propose in Algorithm~\ref{alg:safeDMPC} a secure \dmpc{}, divided into two phases: detection and negotiation.
The new algorithm corresponds to adding a supervision layer for each agent as in Fig.~\ref{fig:safe_dmpc_graph}
\SetKwBlock{negotPhase}{ Negotiation Phase:}{}
\SetKwBlock{detectPhase}{ Detection Phase:}{}
\begin{algorithm2e}[h]
  \DontPrintSemicolon
  \detectPhase{
    Coordinator sends sequence of $\vec{\theta}_{i}^{o}$, $o\in\set{O}$ \;
    Subsystems solve~\eqref{eq:DOP_local}, and send $\tilde{\vec{\lambda}}^{o}_{i}$, $o\in\set{O}$\;
    Coordinator estimates $\widehat{\tilde{P}_{i}^{1}}[k]$ and $\widehat{\tilde{\vec{s}}_{i}^{1}}[k]$ with \EM{}\;
    Coordinator computes $\mathfrak{D}_{i}$ using~\eqref{eq:2}\;
  }
  \negotPhase{
  Apply Algorithm~\ref{alg:quantityAlg} with adequate versions of $\vec{\lambda}_{i}\p$:\;
  \quad$\tilde{\vec{\lambda}}_{i}\p$, if $\mathfrak{D}_{i}=0$ and ${\vec{\lambda}_{i}}_{\mathrm{rec}}$, if ${\mathfrak{D}_{i}=1}$~\eqref{eq:lambda_reconstruction}\;
 }
 \caption{Secure DMPC.}\label{alg:safeDMPC}
\end{algorithm2e}

\section{Example: Temperature Control}\label{sec:App_dMPC}
The system consists of $4$~distinct~rooms (I to IV), which we want to control the air temperatures inside each one of them.
The systems are modeled as continuous-time linear \mbox{time-invariant} systems using the \mbox{3R-2C} model, with parameters in Tables~\ref{tab:modelParamMeaning} and~\ref{tab:modelParam}, and dynamics
\begin{equation}
\begin{matrix}
  \label{eq:systems_cont}
\dot{\vec{x}}_{i}(t)&=&{A_{c}}_{i}\vec{x}_{i}(t) &+& {B_{c}}_{i}\vec{u}_{i}(t)\\
\vec{y}_{i}(t)&=&{C_{c}}_{i}\vec{x}_{i}(t) &&
\end{matrix},
\end{equation}
with
\begin{equation*}
  \label{eq:room_system}
  \begin{matrix}
\resizebox{.98\columnwidth}{!}{$
  A_{\mathrm{c}_{i}}=\left[
    \begin{matrix}
      -\frac{1}{C^{\text{walls}}_{i}R^{\text{oa/ia}}_{i}}-\frac{1}{C^{\text{walls}}_{i}R^{\text{iw/ia}}_{i}}& \frac{1}{C^{\text{walls}}_{i}R^{\text{iw/ia}}_{i}}\\
      \frac{1}{C^{\text{air}}_{i}R^{\text{iw/ia}}_{i}} &-\frac{1}{C^{\text{air}}_{i}Ro_{i}}-\frac{1}{C^{\text{air}}_{i}R^{\text{iw/ia}}_{i}}
    \end{matrix}\right]
    $}
  \\
  \begin{matrix}
    B_{\mathrm{c}_{i}}=\left[
      \begin{matrix}  \frac{10}{C^{\text{walls}}_{i}}& 0\end{matrix}
    \right]\T&C_{\mathrm{c}_{i}}=\left[\begin{matrix}1 & 0\end{matrix}\right]
  \end{matrix}
  \end{matrix},
\end{equation*}
where ${\vec{x}_{i}=[{{x}_{A}}_{i}\T\ {{x}_{W}}_{i}\T]\T}$. ${x_A}_i$ and ${x_W}_i$ are the mean temperatures of the air and walls inside room~$i$. $\vec{u}_{i}$ is the input (the heating power)
for the corresponding room. The inputs are constrained by ${\sum_{i=1}^{4}\vec{u}_{i}(t)\preceq 4\mathrm{kW}}$.

\begin{table}[b]
  \centering
  \caption{Model Parameters}\label{tab:modelParamMeaning}
  \begin{tabular}[b]{cl}
    \toprule
    Symbol&Meaning\\
    \midrule
    $C^{\text{air}}_{i}$&Heat Capacity of Inside Air\\
    $C^{\text{walls}}_{i}$&Heat Capacity of External Walls\\
    $R^{\text{iw/ia}}_{i}$&Resist. Between Inside Air and Inside Walls\\
    $R^{\text{ow/oa}}_{i}$&Resist. Between Outside Air and Outside Walls\\
    $R^{\text{oa/ia}}_{i}$&Resist. Between Inside and Out.\ Air (from windows)\\
    \bottomrule
  \end{tabular}
\end{table}

  \begin{table}[b]
  \centering
  \caption{
    Parameters for each agent}\label{tab:modelParam}
  \begin{tabular}[t]{cccccc} \toprule
    Symbol& I & II & III & IV &Unit\\
    \midrule
    $C^{\text{walls}}$    &$5.4$&$4.9$&$4.7$&$4.7$ &$10^{4}\mathrm{J/K}$ \\
    $C^{\text{air}}$               &$7.5$ &$8.4 $&$8.2$ &$7.7$&$10^{4}\mathrm{J/K}$  \\
    $R^{\text{oa/ia}}$               &$5.2$&$4.6$&$4.9$&$5.4$&$10^{-3}\mathrm{K/W}$ \\
    $R^{\text{iw/ia}}$               &$2.3$&$2.4$&$2.3$&$2.9$&$10^{-4}\mathrm{K/W}$\\
    $R^{\text{ow/oa}}$               &$1.5$&$0.6$&$0.7$&$0.7$& $10^{-4}\mathrm{K/W}$ \\
    \bottomrule
  \end{tabular}
\end{table}

The subsystems are discretized using zero-order hold discretization method with sampling time
${T_{s}=0.25
\mathrm{h}}$
and the quantity decomposition-based \dmpc{} is implemented using prediction horizon ${\predhorz=4
}$.

Three scenarios are simulated for a period of $12.5
\mathrm{h}$:

  \begin{enumerate}
    \item Nominal behavior.
    \item \mbox{Agent I presents non-cooperative behavior} for ${k\geq25}$.
    \item \mbox{Agent I presents non-cooperative behavior} for ${k\geq25}$, with correction.
  \end{enumerate}

  For scenarios (2) and (3), Agent I uses
  $$T_{I}=\left[\begin{smallmatrix}
  14.43288267 & 0. & 0. & 0.\\
  0. & 13.4590903 & 0. & 0.\\
  0. & 0. & 6.93065061 & 0.\\
  0. & 0. & 0. & 3.4447393\\
\end{smallmatrix}
\right].$$

\begin{figure}[h]
  \centering
 \includegraphics[width=.90\columnwidth,trim=0 .3cm 0 .2cm,clip]{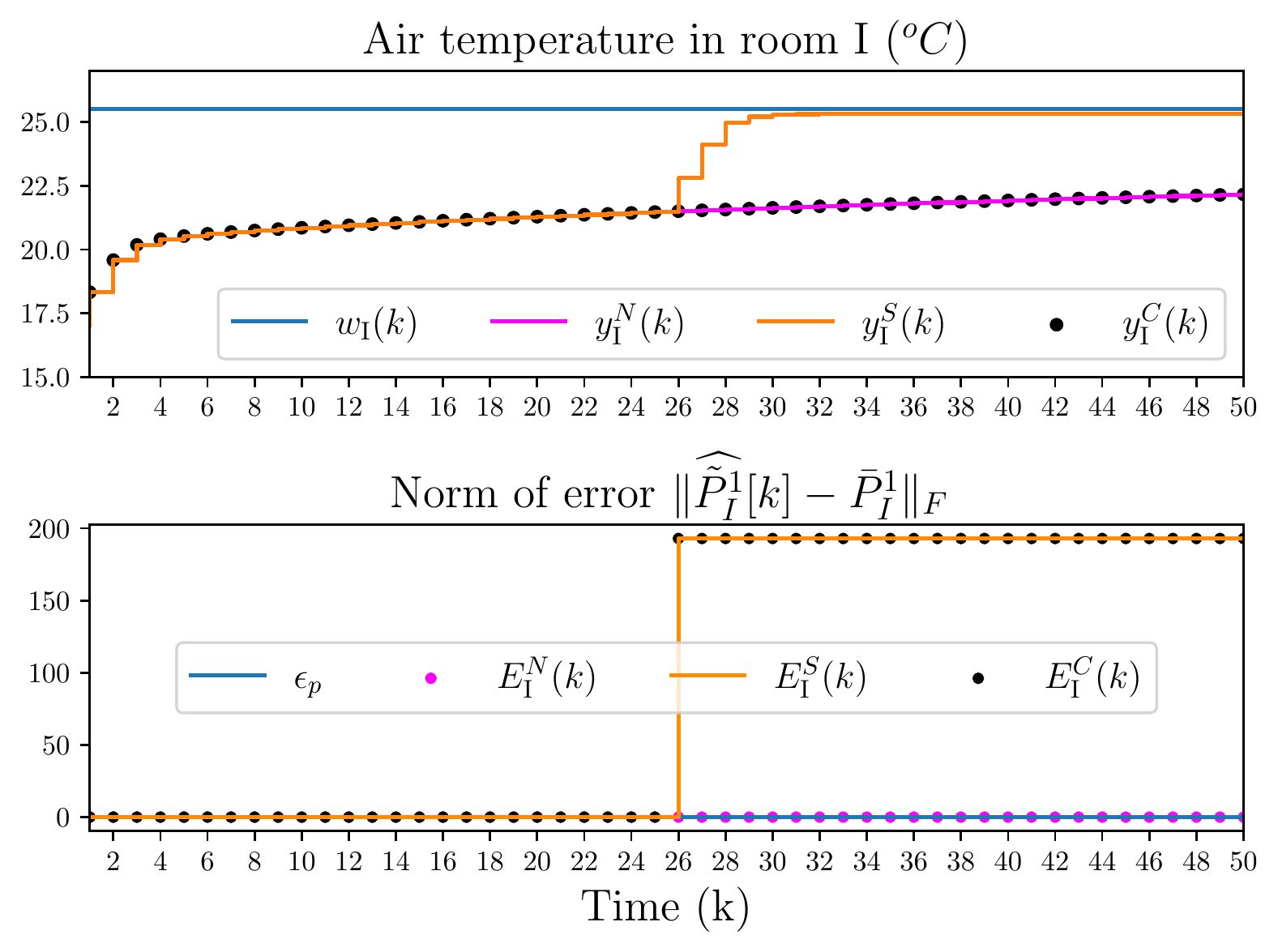}
  \caption{Air temperature in room I and the decision variable $E_{I}[k]$ for three scenarios: nominal (N), selfish behavior (S),
and selfish behavior with correction (C).}\label{fig:response3Scenarios}
\end{figure}

In Fig.~\ref{fig:response3Scenarios}, first, we compare the air temperature in room~I
with its reference (${w_{I}=25.5
^{\circ}}$C), and then the
decision variable ${E_{I}(k)}$ with the threshold $\epsilon_{P}$.
Observe that the reference $w_{I}$ is not reached in the nominal behavior (in magenta),
due to power constraints by which the systems are influenced.
As expected, the decision variable lies under the threshold ${\epsilon_{P}=10^{-4}}$ with values of order ${E_{I}^{N}(k)\approx10^{-10}}$.

When the agent presents a selfish behavior (in orange), the tracking error
${w_{I}-y_{I}}$ is reduced, almost attaining the reference.
In this case, the detection variable surpasses $\epsilon_{P}$,
${E_{I}^{S}\approx200}$, indicating the change of behavior.

\begin{figure}[h]
  \centering
 \includegraphics[width=.90\columnwidth,trim=0 .1cm 0 .3cm,clip]{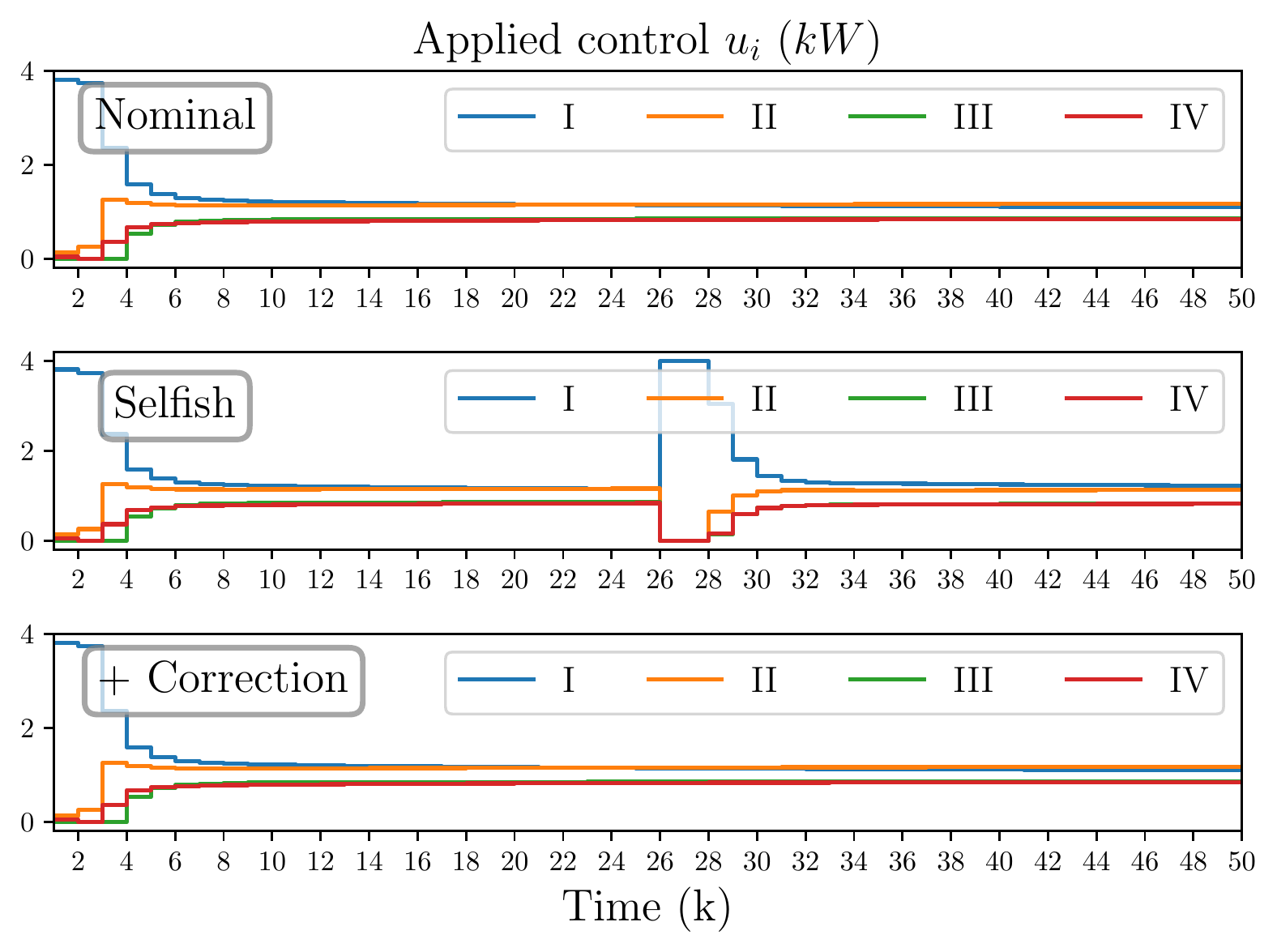}
  \caption{Control applied in all rooms for the 3 scenarios.}\label{fig:control_3Scenarios}
\end{figure}

When the correction is activated in the system, the temperatures approach their nominal value $y_{I}^{N}$.
We can also illustrate the good performances of our proposition by comparing the inputs in Fig.~\ref{fig:control_3Scenarios}.
When room~I is selfish, its control increases while other rooms' decrease. When the correction is activated, it approaches the nominal.

We can also evaluate the performance of the proposed mechanism by comparing the objective functions calculated using the period of simulation ${N=50}$ for the three scenarios (Table~\ref{tab:costsGlobalLocal}).
When agent~I is selfish, we see the decrease in its objective ($\approx\!-40\%$), degrading the overall performance ($\approx\!+10\%$).
When the correction mechanism is activated, the absolute percentual error is ${|\frac{J^{C}-J^{N}}{J^{N}}|\leq10^{-8}}$.
\begin{table}[h]
  \centering
  \caption{Objective functions $J_{i}$ (\% error)}\label{tab:costsGlobalLocal}
  \begin{tabular}[t]{cccc}
    \toprule
    Agent  & Nominal & Selfish & + Correction\\
    \midrule
    I & $ 35008.7 $ ($ 0.0 $)& $ 21969.6 $ ($ -40.0 $)& $ 35008.7 $ ($ -0.0 $)\\
II & $ 29495.3 $ ($ 0.0 $)& $ 38867.4 $ ($ 30.0 $)& $ 29495.4 $ ($ 0.0 $)\\
III & $ 24808.7 $ ($ 0.0 $)& $ 33266.4 $ ($ 30.0 $)& $ 24808.7 $ ($ 0.0 $)\\
IV & $ 23457.8 $ ($ 0.0 $)& $ 31511.0 $ ($ 30.0 $)& $ 23457.8 $ ($ 0.0 $)\\
Global & $ 112770.6 $ ($ 0.0 $)& $ 125614.4 $ ($ 10.0 $)& $ 112770.6 $ ($ -0.0 $)
\\
    \bottomrule
  \end{tabular}
\end{table}

\section{CONCLUSION AND FUTURE WORKS}\label{sec:CC}
In this paper, we proposed an algorithm for monitoring and correcting exchanges between agents in a resource-sharing framework.
The first phase of the algorithm identifies the attacker by exploiting the exchange structure.
After this identification, if necessary, it is possible to reconstruct the original mechanism and recover nominal optimality by inverting the attack.
This principle should be generalized to cases when the attack is not entirely invertible, reconstructing by parts the original mechanism.
Also, other decomposition structures and attack models need to be explored, which we plan to do shortly.

\bibliography{bibliography}
\end{document}